\documentclass[aps,twocolumn]{revtex4}
\usepackage{graphicx,subfigure}
\usepackage{amsmath,amssymb}
\usepackage{bm}

\newcommand{\be}{\begin{eqnarray}}
\newcommand{\ee}{\end{eqnarray}}

\def\slashchar#1{\setbox0=\hbox{$#1$}           
   \dimen0=\wd0                                 
   \setbox1=\hbox{/} \dimen1=\wd1               
  \ifdim\dimen0>\dimen1                        
 \rlap{\hbox to \dimen0{\hfil/\hfil}}      
  #1                                        
 \else                                        
    \rlap{\hbox to \dimen1{\hfil$#1$\hfil}}   
    /                                         
 \fi}                                         %

\begin{document}

\title{Hadronic Correlation Functions in the Random Instanton-dyon Ensemble}

\author{ Rasmus Larsen   and Edward  Shuryak }

\affiliation{Department of Physics and Astronomy, Stony Brook University,
Stony Brook NY 11794-3800, USA}

\begin{abstract}
It is known since 1980's that the instanton-induced 't Hooft effective Lagrangian
not only can solve the so called $U(1)a$ problem, by making the $\eta'$ meson heavy etc,
but it can also lead to chiral symmetry breaking. In 1990's it was demonstrated that,
taken to higher orders, this Lagrangian correctly reproduces effective forces in
a large set of hadronic channels, mesonic and baryonic ones. Recent progress 
in understanding gauge topology at finite temperatures is related with the so called {\em instanton-dyons}, the constituents of the instantons. Some of them, called $L$-dyons,
possess the anti-periodic fermionic zero modes, and thus form a new version of the 
't Hooft effective Lagrangian. This paper is our first study of
a wide set of hadronic correlation function. We found that, at the lowest temperatures at which this approach
is expected to be applicable, those may be well compatible with what is known 
about them based on phenomenological and lattice studies, provided $L$ and $M$
type dyons are strongly correlated.
\end{abstract}
\maketitle

\section{Introduction}

\subsection{Instanton-dyons }
Instantons are the 4-d topological solitons of the (Euclidean) gauge theory,
discovered by Polyakov and collaborators \cite{Belavin:1975fg}. The  so called Instanton Liquid Model (ILM)
has been proposed in \cite{Shuryak:1981ff}. Its main original application was 
related with explanation of chiral symmetry breaking, via collectivization
of the so called Zero Mode Zone (or ZMZ for short). 
Another way to explain it is to state that the hypothetical 4-fermion interaction of the 
Nambu-Iona-Lasinio model \cite{Nambu:1961tp} is in fact the instanton-induced 't Hooft Lagrangian.
One may compare its two
phenomenological parameters -- the mean instanton size $\bar \rho \approx 1/3\, fm$ and  
the total instanton-antiinstanton density $n\approx 1 \, fm^{-4}$ 
-- to two parameters of the NJL model, the coupling constant $G$ and the cutoff $\Lambda$.
Of course, the 't Hooft vertex does more than the NJL operator:  in particular, it
knows about chiral anomaly and correctly breaks the $U(1)_A$ 
symmetry. It also has a natural form factor, allowing to calculate diagrams of any order.

Further development, of the Interacting Instanton Liquid Model (IILM) in 1990's 
has basically included the  't Hooft Lagrangian to all orders. The resulting theory
was shown to reproduce well not only properties associated with the chiral symmetry breaking, the pions and their interactions, but also the correlation functions in
such channels as vector and axial mesons, octet and decuplet baryons, and even glueballs,
for a review see \cite{Schafer:1996wv}. Among shortcomings of this theory
is its inability to describe confinement.

The  deconfinement order parameter, being nonzero at $T>T_c$, is the so called Polyakov line. Its vacuum expectation value $<P(T)>\neq 0$ has been derived in multiple lattice works. 
It is interpreted as the appearance of the nonzero ``holonomy field"  $<A_4(T)>\neq 0$.
Modification of the instanton solution to such environment has lead to the discovery
of the KvBLL caloron solution \cite{Kraan:1998sn,Lee:1998bb} and realization that
instantons can be disassembled into  constituents, now called
instanton-monopoles or instanton-dyons. They are allowed to have non-integer
topological charge because they are connected only by (invisible) Dirac strings.
Since these objects have nonzero electric and magnetic charges and source
Abelian (diagonal) massless gluons, the corresponding ensemble is 
an ``instanton-dyon plasma", with long-range Coulomb-like forces between constituents.  

The first application of the instanton-dyons were made soon after their discovery
in the context of supersymmetric gluodynamics \cite{Davies:1999uw}. This paper solved a puzzling
mismatch  of the value of the gluino condensate, between different answers obtained 
in various approaches. Diakonov and collaborators (for review see \cite{Diakonov:2009jq} )
 emphasized that, unlike the (topologically protected) instantons, the dyons are charged and thus interact directly with
 the Polyakov line. They suggested that since such dyon (anti-dyon) ensemble become denser
at low temperatures, their back reaction  may overcome the perturbative potential and drive it
to its confining value, $< P>\rightarrow 0$.
 A semi-classical  confining regime has been defined by Poppitz et al
~\cite{Poppitz:2011wy,Poppitz:2012sw}   in a carefully devised setting of softly broken supersymmetric models.
 While the setting includes a compactification on a small circle, with  weak coupling and
 an {\em exponentially  small}  density of dyons, the minimum at the confining holonomy
  value is induced by the repulsive interaction in the dyon-antidyon molecules (called  
 $bions$ by these authors). 

Recent progress to be discussed below is related to studies of the instanton-dyon ensembles.
One series of papers were devoted to high-density phase and mean field
approximation \cite{Liu:2015ufa,Liu:2015jsa,Liu:2016thw,Liu:2016mrk,Liu:2016yij}. 
Our efforts were so far focused  on the direct
numerical simulation of the dyon ensembles \cite{Faccioli:2013ja,Larsen:2015vaa,Larsen:2015tso,Larsen:2016fvs} . These works had reproduced
the deconfinement and chiral restoration phase transitions, both in pure gauge ($SU(2)$)
theory and in a QCD-like setting (2 colors and 2 light flavors). They also
show strong modification of both transitions due to unusual quark periodicity phases
\cite{Larsen:2016fvs}.

%
%
Although in this paper we will be using $SU(3)$ color group, for simplicity let us start with
 the simplest case of the $SU(2)$. In the latter case there are only
two selfdual ($L$ and $M$) and two anti-selfdual ($\bar L$ and $\bar M$) instanton-dyon types. Their electric and magnetic charges make all combinations of $\pm 1$. They form three distinct pairs 
$LM,L\bar L,L \bar M$, plus three conjugates, and the amount of screening depends on the 
effective interaction in each of them. 
Two obvious opposite limits are those of weakly correlated or random plasma, for which
the mean field analysis would be adequate. Another limit is very strongly correlated
ensemble. For example, if $LM$ pairs be strongly correlated in their locations, their fields would be
nearly vanishing: in fact this limit would return us to the ``instanton liquid",
in which the solitons are ``neutral", without electric or magnetic charges. Strong correlation in the 
$L\bar L$ channel leads to vanishing magnetic, but not electric fields. 
Strong correlation in the last $L \bar M$ channel would on the contrary cancel electric but not magnetic charges.  

Our previous studies were based on classical effective interaction
derived  from ``streamline configurations" for last two channels  $L\bar L,L \bar M$.
The classical action in the instanton channel $LM$ is different, it is``BPS protected", and so, at the classical level, no interaction was expected (or used in simulations). And yet, as we will show below, 
there are strong phenomenological evidences that even in this channel strong correlations of the instanton-dyons seem to be necessary. 


By the present work we start a set of papers addressing some phenomenologically important issues of the instanton-dyon theory. 

\subsection{ Hadronic correlation functions and structure of the QCD vacuum } 
Two-point correlation functions 
\be K(x-y)=<J(x) J(y)> \ee
of local gauge invariant operators $J$, to be referred as ``currents" for brevity,
are among the most fundamental observables  of QCD. 
Since they are some functions of the distance between the two points $x_\mu-y_\mu$,
 one of the points can always be set to zero.
In Euclidean thermal circle setting, there are two relevant variables, time separation $\tau=x_4-y_4$ and distance $r=\sqrt{\sum_{m=1,2,3}(x_m-y_m)^2}$, and we will systematically put $r\rightarrow 0$, to focus inclusively on their $\tau$ dependence
related to the energy spectrum of the theory.

The correlation functions  are different for operators with different
quantum numbers: 
for a general review of their phenomenology see \cite{Shuryak:1993kg}.
Two-point correlations function, both for mesonic and baryonic operators, 
have been  also has been calculated on the lattice, see e.g.
\cite{Chu:1992mn}, and in the instanton liquid model (see review \cite{Schafer:1996wv}
and references therein).

Let us just remind few key facts.
At 
large distances they decrease exponentially, with the exponent given by
 the ``spectral gap", the lowest excitation in the sector with the corresponding quantum numbers. Their opposite limit of small distances reflects  propagation of the fundamental objects of the QCD, quarks and gluons.
 In between these two limits, one can compare the
 correlation functions to those of free propagation of 
 quarks, and identify ``attractive" and ``repulsive" channels,

Specific combination of the two limits lead to successful parameterizations of the correlation functions,
originally suggested in the context of the QCD sum rules \cite{SVZ}.
The basic relation between the so called ``spectral density", the imaginary part of the Fourier transform of the correlator
in real time, and the real part of the correlator calculated in Euclidean time is given by the dispersion relation.
Its coordinate form is 
\be K(x,T)={\frac{1}{\pi}} \int ds Im \tilde K(s) D(\sqrt{s}, x, T) 
\ee
where the standard Mandelstam's invariant $s=-p^2$ is related with the Minkowskian momentum squared, and $D(M,x,T)$
is the Euclidean propagator of a particle of mass $M$ to Euclidean distance $x$  at temperature $T$.

Out of many possible quantum numbers, corresponding to various mesonic channels, we
selected four most studied ones. Those are all for the  ``charged" isovector channels, say of $\bar u \Gamma d$ flavor structure, which does not require 
(statistically difficult)  disconnected diagrams. The gamma matrices for the pseudoscalar $P$, vector $V$, axial vector $A$ and the scalar $S$ correlators are
\be \Gamma=i \gamma_5,\gamma_0,\gamma_0\gamma_5,1 \ee respectively. The corresponding lowest mesonic states in these channels are
the $\pi^-(134),\rho^-(770),a_1^-(1260)$ and the scalar  $a_0^-(1450)$, the numbers are masses in MeV. 

(In the literature on chiral symmetry breaking the isovector scalar channel 
-- the $U(1)_a$ partner of the pion -- is also known by its old name $\vec\delta$.
Note also that
 the indicated state $a_0^-(1450)$  is the lowest $\bar q q$ meson state with this quantum number. The resonance 
$a_0(980)$ has near-degenerate isoscalar partner $f_0(980)$: those states are believed to be weakly bound  mesonic molecules and thus are disregarded as far as the two-point correlator is concerned.)

Since mesonic masses appear  $squared$ in the effective Lagrangians, consider approximate values of those for the
channels under considerations, $$m^2_{P,V,A,S} \approx 0.02, 0.5, 1.58,2.1\,\,\, GeV^2$$ Two middle ones, the
 vector and axialvector, are the partners under the $SU(N_f)$ chiral symmetry, and their splitting in mass squared  $m^2_{A} -m^2_{V}\approx 1\, GeV^2$ indicate the
 strength of its breaking in the QCD vacuum. 
 (In non-relativistic quark model the vector 
 is  a ``normal" meson, with the mass close to twice the constituent quark mass,
 and the axial vector is the orbital excitation. )  The other two are the $U(1)_a$  chiral symmetry partners,
 and their squared masses are split more, by $m^2_{S} -m^2_{P}\approx 2\, GeV^2$. To accommodate those in the 
non-relativistic quark model one needs to include additional strong attraction/repulsion
of the topological origin. 

While the squared masses give some hints about the scale of $\bar q q$ interaction in these
channels, much more detailed information on that comes from studies
of the corresponding correlation functions.
Theory and phenomenology
of those, first systematically reviewed in \cite{Shuryak:1993kg},  do indeed reveal
 very different $x$-dependence, depending on
the quantum number of $J$. Some channels are ``strongly attractive", with $K(x)$
exceeding the $K_0\sim 1/x^6$ (corresponding to  propagating massless quark and antiquark). 
Some are ``strongly repulsive", while all vector channels ($\rho,\omega,\phi$)  are ``near-free", in the sense that
$K(x)/K_0(x)\approx 1$ in a wide range.  It is those splittings of the correlation functions $K(x)$ 
which we are going to calculate and discuss in this work.

A wider issue related to splittings of these functions is the spin-flavor structure of the nonperturbative
effects in the QCD vacuum, leading to spontaneous breaking of the $SU(N_f)$  and explicit breaking
of the $U(1)_a$ symmetry.  The former issue we will study focusing on the $difference$ between the
vector and axial (isovector) correlation functions, $V-A$ for short. The latter one is related with
the splitting between the pseudoscalar and scalar (isovector) correlation functions, $P-S$.

The $V\pm A$ combinations of the correlation functions are
especially valuable.  First of all,  they
can be deduced directly from experimental data, with good (few percent) accuracy. The vector ones have
the spectral densities directly measurable via reaction $e^+e^-\rightarrow hadrons $.
The axial ones are amenable via weak decays, most prominently of the 
reaction $\tau\rightarrow \nu_\tau+hadrons $. 
ALEPH collaboration data  \cite{Barate:1997hv,Barate:1998uf}
remain the best one,  
used in both instanton study \cite{Schafer:2000rv} and recently in the lattice calculation  Ref.\cite{Tomii:2017cbt}. 

From the theoretical point of view, the best for our purposes is the $difference$
$V-A$ of the vector and axial correlators. Due to chiral symmetry, pQCD diagrams
with any number of exchanged gluons contribute equally to both of them, and are canceled
in the difference. What remains are only the non-perturbative chiral symmetry breaking effects, which
we focus on. We will specifically use $V-A$ combination of correlators below to determine the key
parameter of
the instanton-dyon ensemble.

\begin{figure}[htbp]
\begin{center}
\includegraphics[width=7cm]{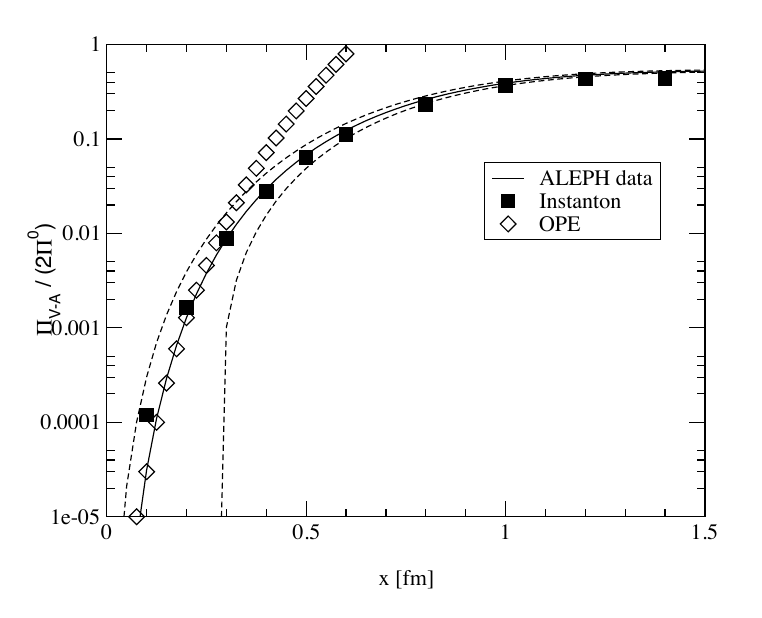}
\includegraphics[width=7cm]{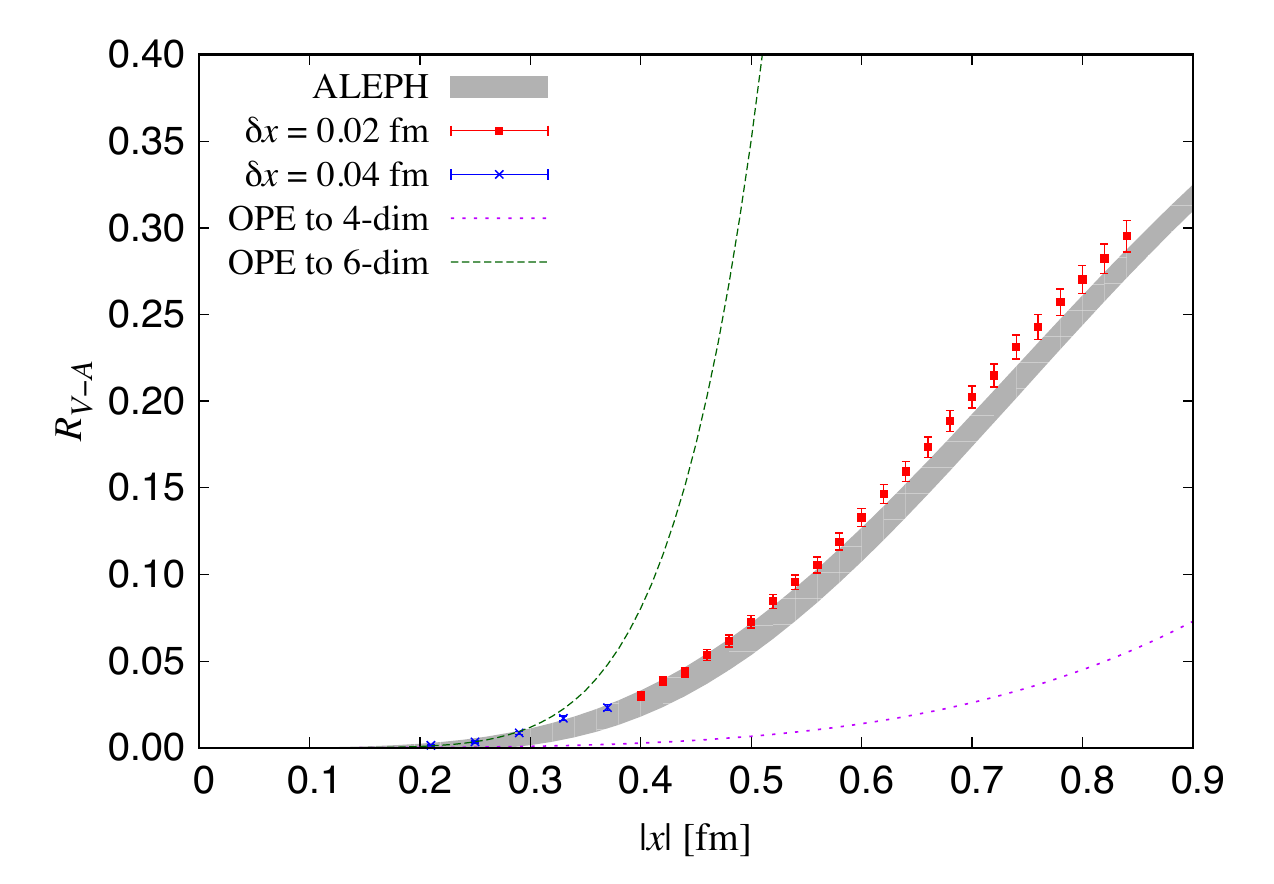}
\caption{ (Color online)  $V-A$ correlation function as a function of the Euclidean distance $x$.
The upper figure, from  \cite{Schafer:2000rv}, compares the ALEPH data 
(region between two dashed lines)
to the instanton liquid calculation (closed squares) and the OPE \cite{SVZ}
(open rhombs). The lower plot, from 
\cite{Tomii:2017cbt}, compares  the same ALEPH data (shaded region) with versions of the OPE (lines) and to extrapolated results of their  lattice simulations (red points). 
}
\label{fig_V-A}
\end{center}
\end{figure}

In Fig. \ref{fig_V-A} we show the $V-A$  combination of correlators deduced from experimental ALEPH results,
the instanton liquid calculation \cite{Schafer:2000rv} (upper plot)
as well as from the recent lattice study \cite{Tomii:2017cbt} (lower plot).
Unlike older studies of point-to-point 
correlators, this one is done with dynamical quarks at physical  mass, with proper continuum extrapolation.
As one can see, both the ALEPH data and modern lattice do provide the correlation function
with the accuracy of just a couple percents. Also it is evident from those plots that
  the original sum rule predictions \cite{SVZ} based on the operator product expansion (OPE)
are applicable  only at very small distances.

The strongest splitting of the correlation function, between the isovector pseudoscalar (charged $\vec\pi$) channel and the scalar
(charged $\vec \delta$), reveals a very important feature of the QCD vacuum/matter structure, namely
its {\em strong inhomogeneity}, but it reveals direct relation to {\em underlying topology} of the gauge fields.
Unfortunately it is
not so accurately known. 

At small $x$ the non-perturbative corrections to correlators -- the splittings -- are approximately
given by expectation values of $<J^2>$, or the $fluctuations$ of the currents in the vacuum.
In a bit more general terms, those are related to VEVs of
various  4-fermion operators.  Strong inhomogeneity of vacuum configurations means that those  fluctuate from point to point   by orders of magnitude. ``Strong"  feature can also be expressed as a statement that some VEVs are large
   $$ <O_{4-fermion}> \,\,\,\gg  \,\,\,  <\bar q q>^2 $$
compared to  the {\em quark condensate squared} in the r.h.s. . 
There are plenty of the 4-fermion operators one can construct out of quark fields, and one may ask
 which ones show this feature in the most pronounced way.
The  studies, in the instanton framework \cite{Schafer:1996wv} and in lattice simulations \cite{Faccioli:2003qz} concluded that it is 
(parts of) the  topology-induced 't Hooft effective Lagrangian. For two light flavors its structure is
\be O_{4-fermion} \sim (\bar u_R \Gamma_i u_L) (\bar d_R \Gamma_i d_L) +(L \leftrightarrow R)\ee
where $L,R$ are left and right handed components of the quark fields and
$\Gamma_i$ may include some color and Dirac matrices. This observation directly implies the
 presence of some small-size topological objects in the vacuum.
Strongly enhanced local violation of $U(1)_a$ chiral symmetry was the key prediction of the 
``instanton liquid model", in which   the typical instanton size is $\rho\sim 1/3\, fm$.
The magnitude of the enhancement is inverse to the ``diluteness fraction" of that model,
the fraction of the 4-volume occupied by instantons $\sim (\rho/R)^4 \sim 1/3^4$.

\subsection{The goals and structure of this paper}

 As we already mentioned above, the modern version of the semiclassical theory 
 at temperatures comparable to the critical one is
 based not on instantons themselves, but on ensemble of their constituents, the instanton-dyons.
Those came into existence due to  inclusion of  the nonzero VEV of the Polyakov line, also known as the ``holonomy
Higgsing".  So far its trust was focused on the deconfinement and chiral phase transitions. 
Now we know that both of them can be reproduced by it, it is time to focus on
the applicability limits of this theory, and see whether it does or does not reproduce
correctly known effects as the theory approaches its boundary. 
 
 Without much details, let us state what is known about the limits of its applicability.
 The upper boundary is expected to be around $T\sim (2-3) T_c\sim 300-400\, MeV$, where the Polyakov line VEV gets trivial $<P>\rightarrow 1$. At higher temperatures
 the $L$-type dyons basically become the instantons themselves, while the $M$ type 
 dyons disappear. 
 
  Our attention in this work is focused on the lowest temperatures at which the instanton-dyon approach 
is  expected to be applicable. It is clear that at $T\rightarrow 0$ it cannot 
be used, with the dyons being basically the 3-d solitons, and with their properties all normalized to $T$.
As we detail below, at low $T$  interference between the dyon fields do lead to
approximately 4-d spherically symmetric instantons, but these interferences are complicated.

  The main question we try to answer in this work is whether the instanton-dyon ensemble can
  correctly reproduce the known features of hadronic correlation functions.
  It would be nice to have lattice data on the correlators as a function of the temperature: yet so far
we only know them quantitatively at $T=0$, in the QCD vacuum. Below we 
  use the instanton-dyon model at its lowest edge, at the temperature  $T=100\, MeV$, and compare the results with vacuum   correlators.  

In section \ref{sec_setting} we briefly outline general properties of dyons in $SU(3)$ and the random ensemble used in this paper. 

In section \ref{sec_zeromodes} we discuss the properties of the fermionic zero modes of the dyons.
For the usual fermionic (anti-periodic in Matsubara time) quarks only one type of $N_c$ dyons, called $L$ dyon, has the fermionic zero modes.  So, naively, in observables related with light quarks, such as the quark condensate, one should only consider sub-ensemble
of $L$ dyons and forget about all $M$. 
However, we will show below that such approach cannot be used,
because in fact those zero modes turn out to be extremely sensitive to $LM$ correlations,
Close proximity of $M$ dyon to $L$ can change local density of the zero modes by up to two
orders of magnitude. 
  We give the formula for the used approximations, the size of the box, amount of $L$ dyons etc. 

In section \ref{sec_corr} we present and discuss our results for the mesonic and baryonic correlation functions. 
We start by showing the sensitivity of the correlators to $LM$ correlations.
We then tune the main parameter  of the model $r_{LM}$ to the 
%
%
best known $V -A$  combination of the correlation functions. After that, we
present  various other correlators. We obviously start with the strongest effect, the $U(1)a$
%
chiral symmetry breaking revealed in the $\vec\pi-\vec\delta$ splitting.
 (The vectors here stand for isovectors of the $N_f=2$
theory.)
At the end we present and discuss the resulting correlators for the  Nucleon and Delta baryonic
currents, and discuss strong attraction in the ``good diquark" channel.


We summarize the paper in section \ref{sec_summary}.





\section{Random Ensemble in SU(3)} \label{sec_setting}
Before we discuss the setting we use for our calculations, it is useful to 
recall the limitations of the semiclassical instanton-dyon theory.

At high $T$ it is limited to the region in which the VEV of the Polyakov line is
not too close to the unit value. The reason for that is that when the holonomy parameter $\nu$ is small,
the $M$-type dyons become too light (and too large) to keep their semiclassical theory meaningful. In QCD
this range approximately correspond to
 $T< 350\, MeV$.

At low $T<T_c$, in the confined and chirally broken hadronic phase,  the holonomy parameter $\nu$ is fixed
to the confining value, so that $<P>=0$ and all types of dyons have the same actions.
 Yet as one moves toward the lower temperatures, the action per dyon $S\sim 1/g^2(T)$ logarithmically  decreases 
due to the running coupling, eventually making their semiclassical theory 
inapplicable. Tentatively we use as the lower ``large" value $S_{M,L}\sim 3\hbar$.
In QCD
this range approximately correspond to $T> 100\, MeV$

(Note that coincidentally this range  correspond well to the temperature range of excited matter
produced in heavy ion collisions at RHIC and LHC colliders.)

In this first paper devoted to hadronic correlation functions we decided to use the simplest
ensemble possible, in which positions and color orientations of the dyons are selected randomly.
We will  thus refer to this ensemble as Random Instanton-Dyon Model, or RIDM.

\begin{figure}[htbp]
\begin{center}
\includegraphics[width=8cm]{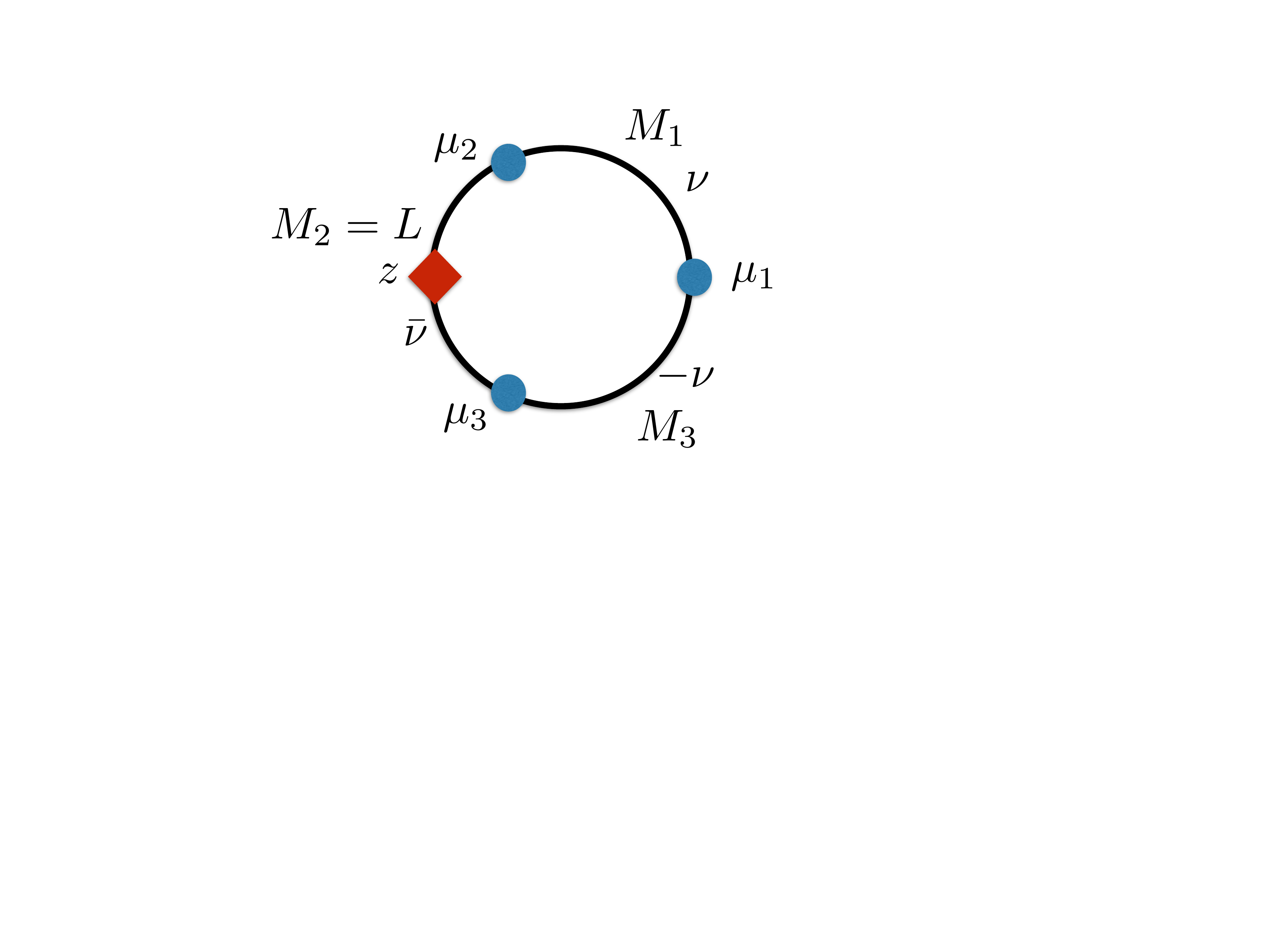}
\caption{ (Color online)  The so called holonomy circle explaining notations used. Three blue points marked $\mu_1,\mu_2,\mu_3$ correspond to three eigenvalues of the Polyakov line in $SU(3)$ 
gauge theory.
Three sectors between those, marked $M_1,M_2=L, M_3$ correspond to actions of the
three types of the dyons. The red rhomb marked $z$ corresponds to periodicity phase
of the antiperiodic quarks: the sector containing it is the one which has the fermionic zero mode.
}
\label{default}
\end{center}
\end{figure}

While our previous works used the simplest $SU(2)$ color group, we now switch to the $SU(3)$.
Therefore we should start by defining the  holonomy parameterization used. 
Standard holonomy phases
$\mu_i, i=1..N_c$ should satisfy the zero-trace condition $\sum \mu_i=0$. In addition, we assume that
$<P>$ is real. These two conditions reduce 3 phases to one free parameter 
\be 
\mu_1=0,\,\,\,\, \mu_2=-\mu_3=\nu
\ee

in terms of which VEV of the Polyakov line is
\be 
<P>= {\frac{1}{3}}+ {\frac{2}{3}} cos(2\pi \nu) 
\ee
The confining value, at which it vanishes, is thus $\nu=1/3$. 

With this definition the actions of $M_i,i=1,2$ dyons are $S_M=S_0 \nu$, where the instanton action $S_)=8\pi^2/g^2(T)$. The action of the ``twisted" $L$-dyon
is $S_L=S_0(1-2\nu)$. In the confining phase all of them have the same action $S_0/3$.

\section{Fermionic zero-modes for correlated $L-M$ dyons} \label{sec_zeromodes}
Zero-eigenstates of the Dirac operator 
play
the central role in our calculation,  as they provide the basic set of wave functions for the region in eigenvalues called 
the Zero Mode Zone (or ZMZ for short)   inside which the long-distance part
of quark propagators is calculated. So, before we embark on modeling
quark propagators and hadronic correlation functions, a direct comparison between those
would be instructive.

We will subsequently discuss three historic approximations:\\
(i) the original periodic instanton (caloron) at $zero$ holonomy \\
(ii) the single instanton-dyon (of the $L$ type)\\
(iii) the KvBLL caloron at $nonzero$ holonomy, or the case of 
a set of interfering $L,M_i$ dyons

 Furthermore, since in the ensemble of the instanton-dyons there are both selfdual and anti-selfdual objects, there are no general formulae for these influences anyway. The  
 practical solution we therefore use is to take as a basis the zero modes of the individual dyons.
 
 The detailed derivation of those  has been done
 in appendix of \cite{Shuryak:2013tka}. Since it was done for arbitrary periodicity phase,
 it includes discussion of the delocalization of zero modes, at the values when
 color holonomy and flavor holonomy values coincide. Here we only need
the zero mode for the physical antiperiodic quark fields, which for
the $L$-type dyons  has the following form
\begin{eqnarray}
\phi _a ^A &=& e^{i \pi t}\sqrt{\frac{\bar{v}^3}{2\pi}}\frac{\tanh (\bar{v} r /2)}{\sqrt{\bar{v}r \sinh (\bar{v}r)}}\epsilon _{aA}     \label{mode_L}
\end{eqnarray} 
Here and below in this section we write everything in units  $T=1$. 
The normalization constant corresponds to $\int d^4 x Tr[\phi \phi^*]=1$.

  The quark zero modes for the  finite-$T$ instantons, known as ``calorons" ,
  are also known.  
 Their gauge potential belongs to a general ansatz
 \begin{eqnarray}
   A_\mu^a &=& \overline \eta^a_{\mu\nu} \Pi(x)\partial_\nu\Pi^{-1}(x)
  \end{eqnarray} 
  which in this case take the form
  \begin{eqnarray}
 \Pi(x) &=& 1 + \frac{\pi\rho^2 T}{ r} \frac{\sinh (2\pi T r)}
 {\cosh (2\pi T r) - \cos (2\pi T \tau) }
\end{eqnarray}
  where $\rho$ denotes the size of the instanton. Note that the dependence on Euclidean time
  $\tau$ is periodic, with the correct period $1/T$.
  
  The fermion zero mode is also expressed in terms of this function
  \begin{eqnarray}  \label{mode_C}
  \label{zm_T}
\psi_i^a &=& \frac{1}{2\sqrt{2}\pi\rho} \sqrt{\Pi(x)}\partial_\mu \left(
  \frac{\Phi(x)}{\Pi(x)}\right) \left(\frac{1-\gamma_5}{2}\gamma_\mu
  \right)_{ij} \epsilon_{aj}\, ,
  \end{eqnarray} 
  where $$\Phi(x)=(\Pi(x)-1)\frac{\cos(\pi T \tau ) }{ \cosh(\pi T r)}$$

Before we are going to compare these functions in more detail,
it is instructive to compare their asymptotic behavior at large $r$. 
Both 
 decay exponentially at large distances, but with different exponents.
 The L-dyon mode (\ref{mode_L}) decreases as 
   $exp(-\bar{v} r/2)$  prescribed by the magnitude of the corresponding holonomy. 
   The caloron zero-mode  (\ref{mode_C}) exponential decay is
$\exp(-\pi r)$ , 
related to the lowest fermionic  Matsubara
frequency. The two match only at high $T$ where
 $\bar{v}\rightarrow 2\pi$. 

 So, already from the comparison of the asymptotic of these modes, one    can preview
    the generic phenomenon: the sizes of the instanton-dyons in general (and their zero modes
   in particular) are
    $larger$ than those of the calorons.  This statement may appear very counter-intuitive,
   since the  instanton-dyons are the caloron constituents. 
   Note however, that interference of the  instanton-dyon fields is mostly $destructive$.
   


 Note also, that since at the higher-T limit  of the 
instanton-dyon theory $\bar \nu\rightarrow 1$, in this limit the zero mode asymptotics
for the $L$-dyon and the instanton match.

The size of the zero mode is also determined by the size parameter of the caloron $\rho$. At high
$T>T_c$ temperatures the instanton density is corrected by the so called Pisarski-Yaffe
factor  which with a good accuracy  is just the result of electric
Debye screening by quarks and gluons scattering on the caloron
\begin{eqnarray} \label{eqn_rho_T}
n_{inst}(T,\rho)&=&n_{inst}(0,\rho) e^{-(\frac{\rho }{ \rho_T})^2} \\ 
\frac{1 }{ \rho_T^2}&=&\frac{2 N_c+N_f }{ 3} \pi^2 T^2 \nonumber
\end{eqnarray} 
which forces the instanton sizes to scale with temperature like $\rho \sim \rho_T\sim 1/T$.
(For explanations see e.g. review {Schafer:1996wv}.)
However at small $T\rightarrow 0$ the instanton sizes have a constant limit,  in the original instanton liquid model $\rho(T\ll T_c) \approx 1/3 \, fm$.
For
 for QCD ($N_c=N_f=3$) 
$\rho_T$ reaches this value at $T=T_{min}\approx 100 \, MeV$.

\begin{figure}[htbp]
\begin{center}
\includegraphics[width=8cm]{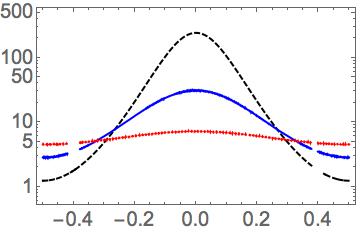}
\caption{ (Color online)  The dependence of the caloron zero mode density on time, for $\rho=\rho_T$ (black dashed),  $\rho=2\rho_T$ (blue solid), and $\rho=1$ (red dotted) lines.
The quantity $\rho_T$ is
defined in 
(\ref{eqn_rho_T}). }
\label{fig_t_dependence}
\end{center}
\end{figure}

The main difference between the zero mode of the caloron and
the L-dyon is that the former is strongly time-dependent.
Substituting  three different values of the size we see, from Figure \ref{fig_t_dependence},
that  the density of the zero mode as a function of time changes by up to  two orders of magnitude when $\rho=\rho_T$, but is weakly time dependent when $\rho=1$ (that is $1/T$
in absolute units).

\begin{figure}[htbp]
\begin{center}
\includegraphics[width=8cm]{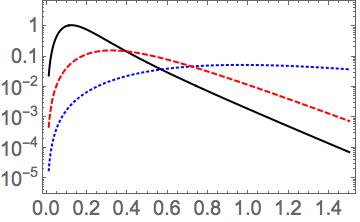}
\includegraphics[width=8cm]{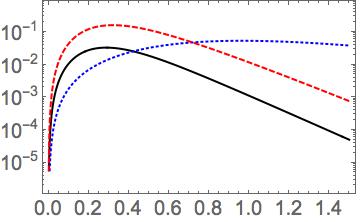}
\caption{ (Color online)  The  dependence of the zero mode densities times $r^2$ on $r$, for $t=0$ (upper plot)
and $t=1/2$ (lower plot). Mote that these times correspond to the maximum and minimum
in the previous plot.
In all of them black solid curve is for the caloron with  $\rho=\rho_T$,
while blue dotted and red dashed curves are for the $L$-dyon, with $\bar{v}=2\pi/3$ and $2\pi$
respectively. 
}
\label{fig_r_dependence}
\end{center}
\end{figure}

The space dependence of the  caloron zero mode density is shown in Fig. \ref{fig_r_dependence}. Here we compare the integrand of the normalization condition,
thus multiply the densities by $r^2$. Note further that there are two dyon curves, corresponding to confining value $\bar{v}=2\pi/3$ (valid at $T<T_c$) and the ``trivial holonomy" value $\bar{v}=2\pi$  valid at high $T$.  Comparison of the plots indicate that
while the ensemble of the calorons can be relatively dilute, that of the dyons cannot be
such, because their zero modes have significantly larger range in space.

A popular measure of how strongly the function is localized is the integral of the density $squared$, or the 4-th power of the zero mode
\be 
I_4=\int d^4 x [Tr(\phi \phi^+)]^2
\ee
(Let us remind that the integral of the second power is the normalization integral taken to be 1, and that
the 4-fermion operator -- 't Hooft effective Lagrangian -- is instrumental in breaking
the chiral symmetry. )

For the caloron radii $\rho=\rho_T, 2\rho_T,1$ (the same as shown in Fig. \ref{fig_t_dependence}) its values are $I_4=26.6, 4.20, 1.55$, respectively. 

All these comparisons suggest the inevitable conclusion: 
in distinction to the ``instanton liquid" -- which 
is relatively dilute, with the instantons occupying only few percent of the volume
\cite{Shuryak:1981ff} -- the ensemble of the instanton-dyons at $T<T_c$ is 
in fact rather dense. 

Finally, we discuss the fermionic zero mode for a KvBLL caloron at $nonzero$ holonomy  worked out
by van Baal and collaborators \cite{Bruckmann:2003ag}, using general ADHM and Nahm construction.  That  
 resulted in very complicated expressions which we do not to copy here.
 The effect we are after takes into account mutual influence of the fields of $L$ and $M$
 instanton-dyons, as a function of their relative distance, related to the ``caloron size" parameter $\rho$ via
 \be r_{LM}=| \vec r_L -\vec r_M |=\pi \rho^2 T\ee
 
 \begin{figure}[h!]
\begin{center}
\includegraphics[width=8cm]{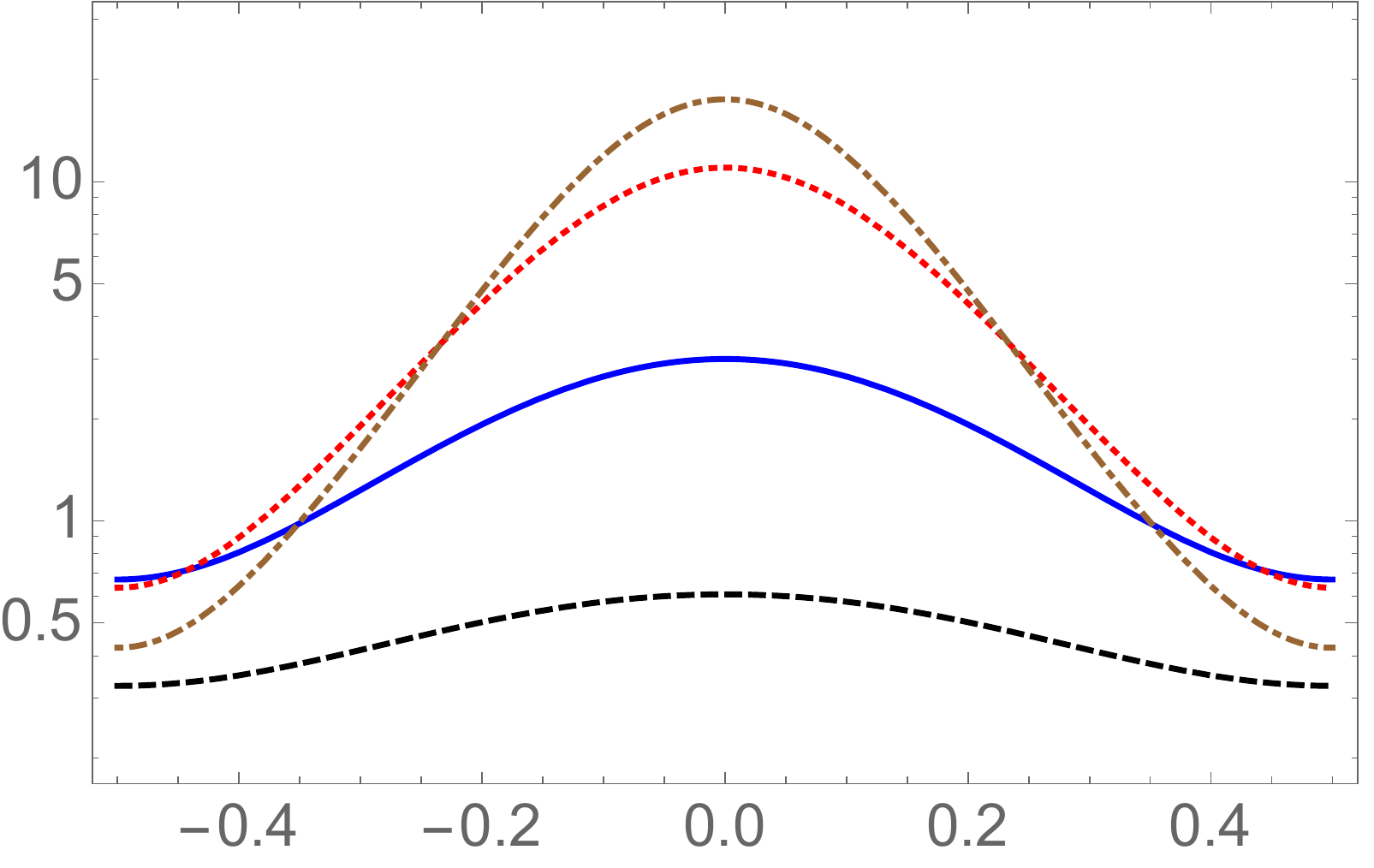}
\caption{ (Color online)  The  time dependence 
of the zero mode densities, at $r=0$, for the SU(2) caloron at
confining holonomy $v=\bar{v}=\pi$.
The lowest (black dashed) curve 
is at relative distance 1, the next (blue solid) is 0.5,  then (red dots) 0.2 and 
(brown dash-dotted) one 0.1.  The time and distances are in units such that $\beta=1/T=1$.
}
\label{fig_t_vanBaal}
\end{center}
\end{figure}

 The results are shown in Fig. \ref{fig_t_vanBaal}: one can see that if the distance between the dyons is as large as 1 (the lowest black dashed curve), the time dependence is rather mild,
 resembling the infinite distance (single dyon) case discussed above, in which there is $no$
 t-dependence at all. But, as  $L$ and $M$ are moved closer to each other, their interference
 deform the zero mode to be well localized. Indeed, close  $L-M$ dyon pair is a small dipole,
 with electric and magnetic fields canceling outside. So the fermionic zero mode
 get strongly localized in between them.

The density of the zero modes can be written in a nice form
(see e.g. (11) of   \cite{Bruckmann:2003ag})
\be \hat \Psi^a_z(x) ^+ \hat \Psi^b_z(x)= - \frac{1 }{ (2\pi)^2} \partial_\mu^2 \hat f^{ab}_x(z,z) 
\label{eqn_f}
\ee
 where the r.h.s. is the Green function of certain equation in Nahm variable $z$.

\subsection{The gauge factors of the zero-modes}

Since we treat the dyons as individual object and don't include overlap effects, the shape of the dyon is in its basic principle am $SU(2)$ object. Higher order groups are obtained my taking the $SU(2)$ object and injecting it into a higher group, which in this case is the $SU(3)$.

To have more than one dyon in the same gauge, the hedgehog gauge dyon is rotated into a specific direction in color space. As in earlier work we choose to rotate the dyons into the $\tau_3$ direction. In order to do this we first rotate all directions by an angle of $\phi$ around the $\tau _3$, followed by a rotation of $\theta$ or $\pi - \theta$ for dyons and antidyons around the $\tau _2$ direction, putting the direction along the $z-axis$ corresponding to the $\tau _3$ axis. Since any rotation around the $z-axis$ in the $xy$ plane will be invariant, we have a free rotation, corresponding to the $U(1)$ rotation. This angle sets the angle of the core and is important when dyons overlap each other. We therefore use the time coordinate for this rotation.  

\section{The settings}
In our previous simulations
\cite{Larsen:2015vaa,Larsen:2015tso}  in the partition function the classical and one loop interactions
of all dyon pair channels were included.   The color group was $SU(2)$, and the
3-d manifold on which simulation was done was the 3d sphere $S^3$. 

The instanton-dyons we use in this paper
are  embedded in $SU(3)$ color group. It has $L,M_1,M_2$ and their anti-solitons,
 6 species in total
 . The 4-d manifold is the standard periodic  Matsubara  box, with variable space and time dimensions.
 The number of the dyons in  the simulation
we keep constant, $N_i=100$, where $i$ can be $L$ or $\bar{L}$. 

 Since we only consider antiperiodic (fermionic) quarks, only the $L,\bar L$
dyons have quark zero modes. Thus the total basis of the zero mode zone is $N_L+N_{\bar L}=200$ states. The propagation of quarks from one object to another
is done via the ``hopping matrix" $T_{ij}$, in this case the matrix of  $200\times 200$ size.  
 Other dyons $M_i$ only enter via 
their correlation/overlaps with $L,\bar L$, which we describe approximately via the parameter
$r_{LM} $ as detailed below.

The temperature has been set by the size of the box in temporal direction, which was chosen to be $2fm$, while the size of spatial directions was used to control the density. The density was found by fitting to experimental data as shown in section \ref{sec_corr}.


The full zero-mode in $SU(2)$ and $SU(3)$ are known, but it is a huge expression which, even after long simplifications in Mathematica, is not viable to write in reasonably compact form. 
 since its specific form requires derivatives that makes it extremely long. We have therefore 
 generated numerically a set of graphs for their density distribution in space-time $x$ and 
 parameterized those approximately. 
 
 The zero-mode is a function of position $x$, holonomy $\nu$ and distance to the $M$ dyon $r_{LM}$. We were interested in the shape for this at $\nu=1/3$ and for distances $r_{LM}<2$ for which the approximation works reasonably fine.

The form is
\begin{eqnarray}
\psi _A ^a &=& N\sqrt{f}\epsilon _{a A}\\
f &\approx& \exp[(0.4 + \pi \exp[-4r_{LM}])(\cos[2\pi t] - 1)] \nonumber \\
 &\times & 1/\cosh[-(\pi +(2\cos[\theta] + 10)\exp[-2 r_{LM}]) \nonumber \\
 & &   \sqrt{x^2 + y^2 + (z + 0.4 r_{LM}^2)^2]} 2\nu T]\nonumber 
\end{eqnarray}   
Where $N$ is a normalization factor, since the parameterization normalization was was slight of $1$, $\theta$ is the angle between the $L$ dyon to the $M$ dyon and the position of the field. The color structure is given by the $\epsilon $ symbol.

  The so called hopping matrix is made of overlap matrix elements of the Dirac operator, symbolically
\be   T_{ij}=<i | \hat D_\mu \gamma_\mu | j> \ee
Here $D_\mu$ is the covariant derivative including the gauge field. If the fields is just a sum
of fields for each dyon, one can use Dirac equation of the zero mode to remove all fields and
keep only the usual derivative. Those integrals were also
numerically calculated and parameterized as follows

\begin{eqnarray}
T_{ij}&\approx&  \exp[-\sqrt{0.7 + (2\pi(\nu/2 + \bar{\nu}/2.5\exp[-0.5r_{LM}^2])r)^2}]\nonumber\\
 && (1 + 
   Cos[2\pi t]\exp[-7r_{LM}]) 8\pi T \nu
\end{eqnarray}
This parameterization only works for values of $r_{LM}<2$.


The parameterizations made for $SU(2)$ is then embedded into $SU(3)$ and the random ensemble is generated for 200 different $L$ dyons. The size of the box and the constant $r_{LM}$ is varied, until we obtain a difference in the Axial and Vector channel that is similar to the experimental results. The results of this fit is shown in section \ref{sec_corr}. 

The crucial parameter here and in the previous expression is $r_{LM}$, in $SU(2)$
the distance between the $L$ and $M$ dyons.  Its value used will be explained in the
next section.

The correlation functions are given by the Feynman diagrams, in which quark-antiquark pair
for mesons, or three quarks for baryons, propagate from $x$ to $y$. For the quark propagator
we use the approximation well developed for the instantons. Its zero mode part
has the structure $< x | i > (T^{-1})_{ij} < j | y> $ where  $< j | y >=\psi_0(y)$ is the zero mode
of the $j$-the dyon. Note that the propagator includes the $inverse$ hopping matrix,
since the propagator is inverse to the Dirac operator.

For any configuration of the dyons, the set of zero modes are calculated
The  200$\times$ 200 hopping matrix is filled and inverted. The obtained quark propagator
is inserted in all diagrams, convoluted with various matrices in the currrents.
The ensemble of configurations we us include 10 configurations.

\section{The correlators} \label{sec_corr}

\subsection{Mesonic Correlators}

As explained in the Introduction, the  difference in the correlation functions for the vector (charged $\rho$)
and axial (charged $A_1$) channels $V(x)-A(x)$ , related to $SU(N_f)$ chiral symmetry breaking, is the most accurately known topology-related combination, both from the experimental inputs and from
the lattice. However, it is not the difference corresponding to the largest splitting, which is that between the pseudoscalar and scalar channels.

\begin{figure}[htbp]
\begin{center}
\includegraphics[width=8cm]{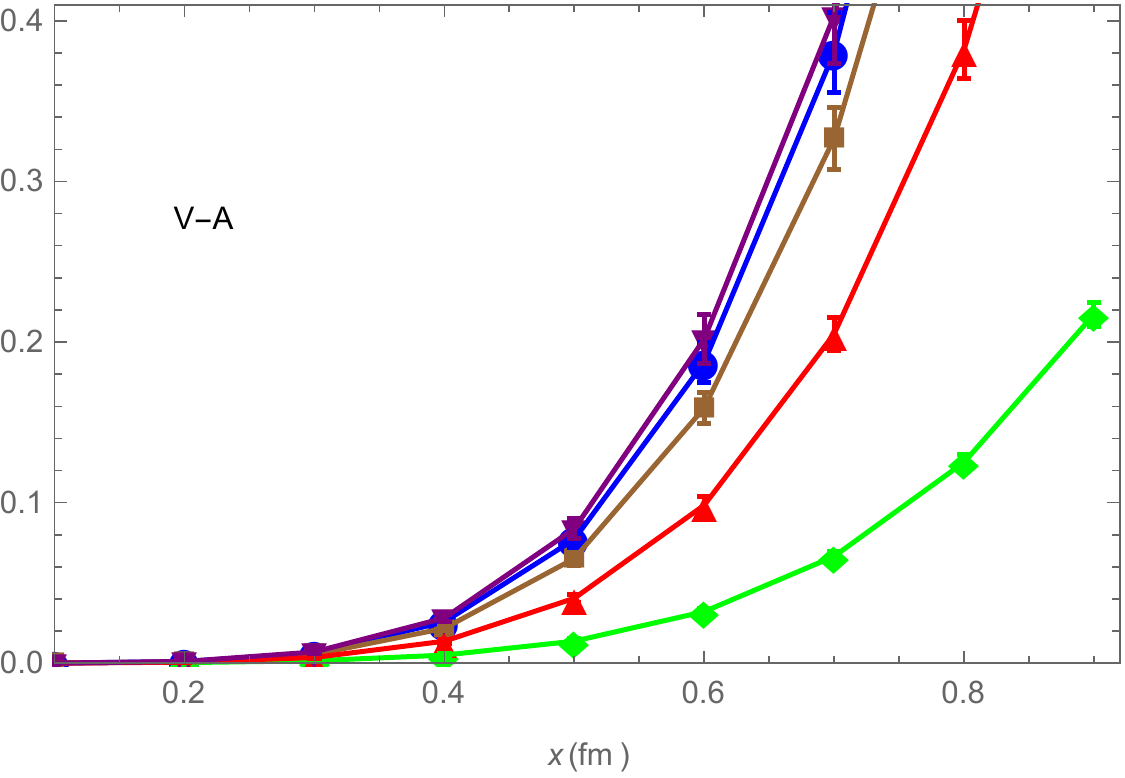}
\includegraphics[width=8cm]{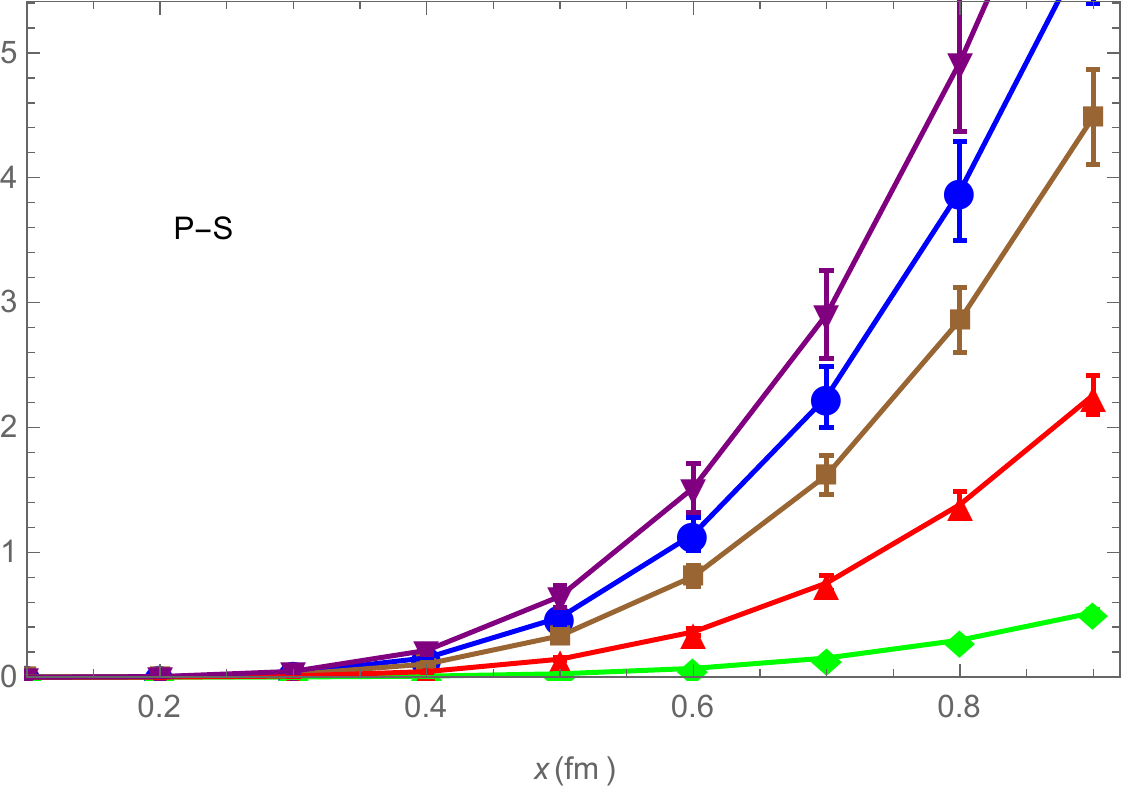}
\caption{ (Color online)   The upper plot is the difference between  the vector and axial correlators $(V(x)-A(x))/(2K_0(x)) $ versus the temporal distance
$x$ (in $fm$). The lower plot is the difference between  the pseudoscalar and scalar correlators  $(P(x)-S(x))/(2K_0(x)) $.
Different curves are for values of the parameter
 $r_{LM} = 0.1(Purple\triangledown )$, $0.15(Blue \bullet)$, $0.2(Orange \Box)$, $0.3(Red \triangle)$, $0.5(Green \diamond)$.}
\label{fig_VA}
\end{center}
\end{figure}

Let us start to display  our results by showing, in Fig. \ref{fig_VA}, how both the $V(x)-A(x)$ and $P(x)-S(x)$ differences depend on the parameter  
$r_{LM}$ of our model.

The first observation from this figure is that indeed the second $P-S$ splitting is much larger than the first $V-A$, by about one order of magnitude. 

\begin{figure}[htbp]
\begin{center}
\includegraphics[width=8cm]{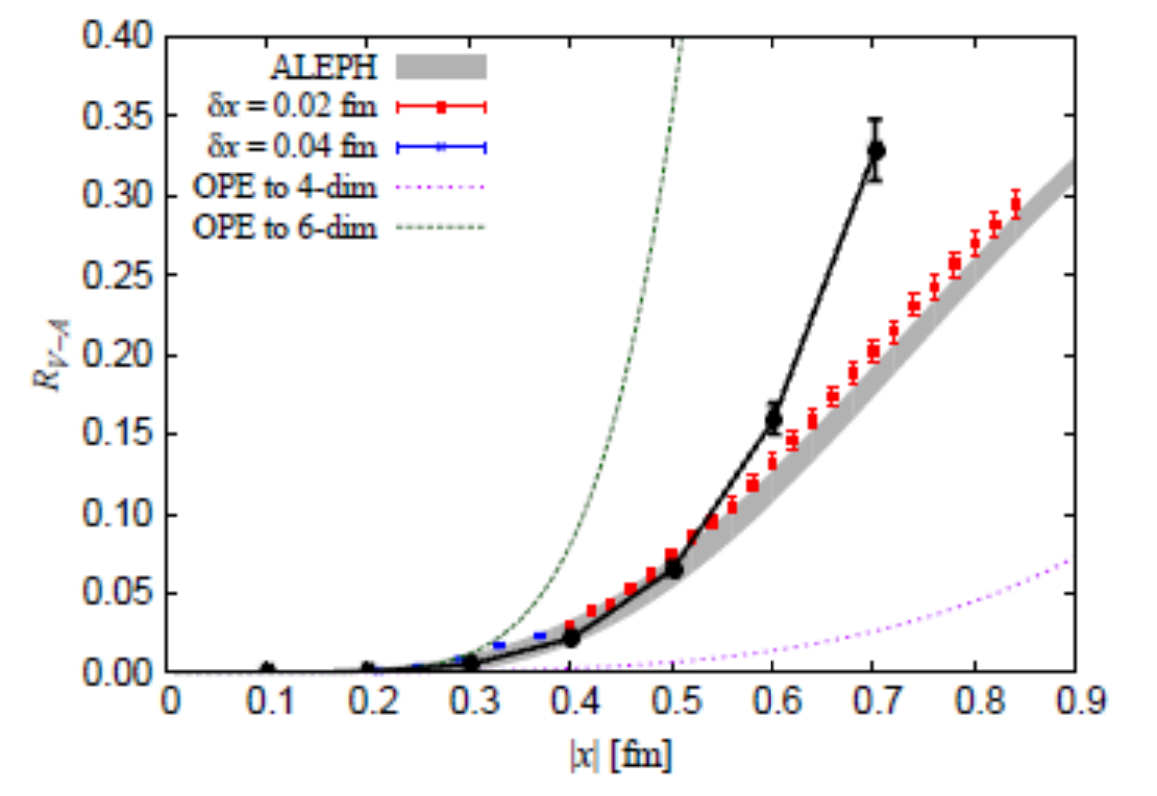}
\caption{ (Color online)  The normalized vector minus axial vector difference 
$(V(x)-A(x))/(2K_0(x)) $
 channels versus the distance $x$ $(fm)$. The narrow shadowed region corresponds to 
ALEPH data, the red and blue dots correspond to the lattice data   \cite{Tomii:2017cbt}, for two lattice spacings indicated on the plot. 
 Our results for $r_{12}=0.2$ are shown by (black) $\bullet)$.}
\label{fig_overlap}
\end{center}
\end{figure}

The second observation,  clearly seen in both of the plots, is that they are quite sensitive to the magnitude of the 
parameter $r_{LM}$, the typical distance from the $L$ dyon to the $M$ dyons. By
 varying it one finds very different magnitude of the correlation functions. 
Therefore, using this sensitivity 
 we can tune the value of this parameter to correspond to the known vacuum
value of the $V-A$ correlator, see Fig. \ref{fig_overlap}. The used value for the fit were $r_{12}=0.2$.

We see that the fit works will up to distance about $0.5\, fm$, but after this overshoots the experimental and lattice data at $|x|>0.5 \, fm$. In to the latter region one also observes several unphysical
effects, in particular the scalar correlator gets negative $S(x)<0$, see 
Fig. \ref{fig_overlap},
in contradiction to spectral decomposition which require all diagonal correlation functions to be strictly positive.

These abnormal phenomena in fact has been observed long before, in random instanton liquid model
(RILM) and later in quenched QCD  \cite{Chu:1992mn}. Note that both of  these approaches
lack the fermionic determinant in the measure, and thus lack the most critical 
back reaction of quarks on the topological ensemble. Arbitrary operations like ``quenching"
break connections between these ensembles and quantum field theory foundations,
so the correlator positivity and other general features of QFTs can and are violated.

 It has been later shown (see review {Schafer:1996wv}) that in the so called interacting instanton liquid model (IILM) -- which includes 
the fermionic determinant in the measure --  these abnormal phenomena disappear.
And they, of course, also are not present in unquenched lattice simulations with the dynamical quarks.
So, although we have not yet done simulations with fully interacting (unquenched) 
ensemble fo SU(3) instanton-dyons, we are confident that in this case these abnormalities
would disappear as well.



%

\begin{figure}[htbp]
\begin{center}
\includegraphics[width=8cm]{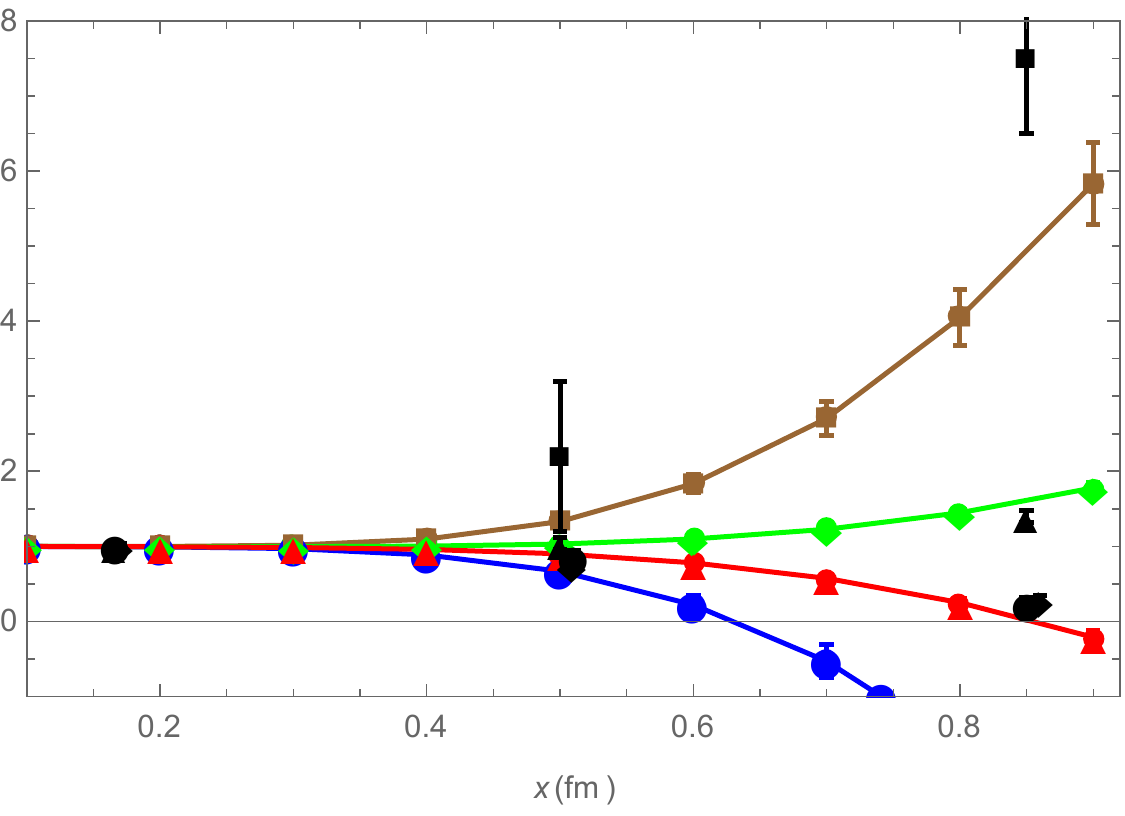}
\caption{ (Color online)  The colored points connected by lines are our results for four channels,
 for $r_{12} = 0.2$. 
Top to bottom: Pseudoscalar (Brown) $\Box$,  Vector (Green) $\diamond$ , Axial vector
(Red) $\triangle$, and
Scalar (Blue) $\bullet$. 
  The individual (black) points  without lines are lattice data from \cite{Chu:1992mn}, their symbols are the same as for our data.}
\label{fig_all_4}
\end{center}
\end{figure}




Now we return to Fig. \ref{fig_all_4} in which the correlations functions are shown
for all four channels  under consideration, $P,V,A,S$ from top to bottom.
 One can clearly see, that for small distances $x< 1/3 \, fm$
all of them are in a good approximation identical. We further remind that their value in this region, equal to $1$ in our normalization, corresponds to
 free propagation of the massless quark and antiquark.
 
 At larger distances in Fig.\ref{fig_all_4}  our simulations for the four channels display clear splitting pattern, which is nearly identical
 to what was first observed in RILM and then on the lattice in 1990's.
 The lines go upward correspond to attractive channels $P,V$ and those going downward 
 show repulsion in the $A,S$ channels.
For comparison we also show in this figure the results from \cite{Chu:1992mn},
shown by similar symbols as ours but without connecting lines.
   Overall our results are reasonably well consistent with  these lattice data.
   On a quantitative level one finds certain differences: e.g. the splitting of our pseudoscaler is slightly weaker than on the lattice. All these differences are however completely
   understandable and are due to different values of the quark masses in our ensemble
  and on the lattice. 

Last subject we would like to discuss for the mesonic correlators is how 
they change as the temperature increases. 
These changes are supposed to be caused by (at least) the following effects:\\
(i)  the VEV of the Polyakov line moves toward trivial value 1, and thus the holonomy 
parameter $\nu$ goes towards 0;\\
 (ii) the effective coupling runs to smaller values, the action of the dyons grow and their
 density  decreases;\\
(iii) the size of the Matsubara box decreases\\

\begin{figure}[htbp]
\begin{center}
\includegraphics[width=8cm]{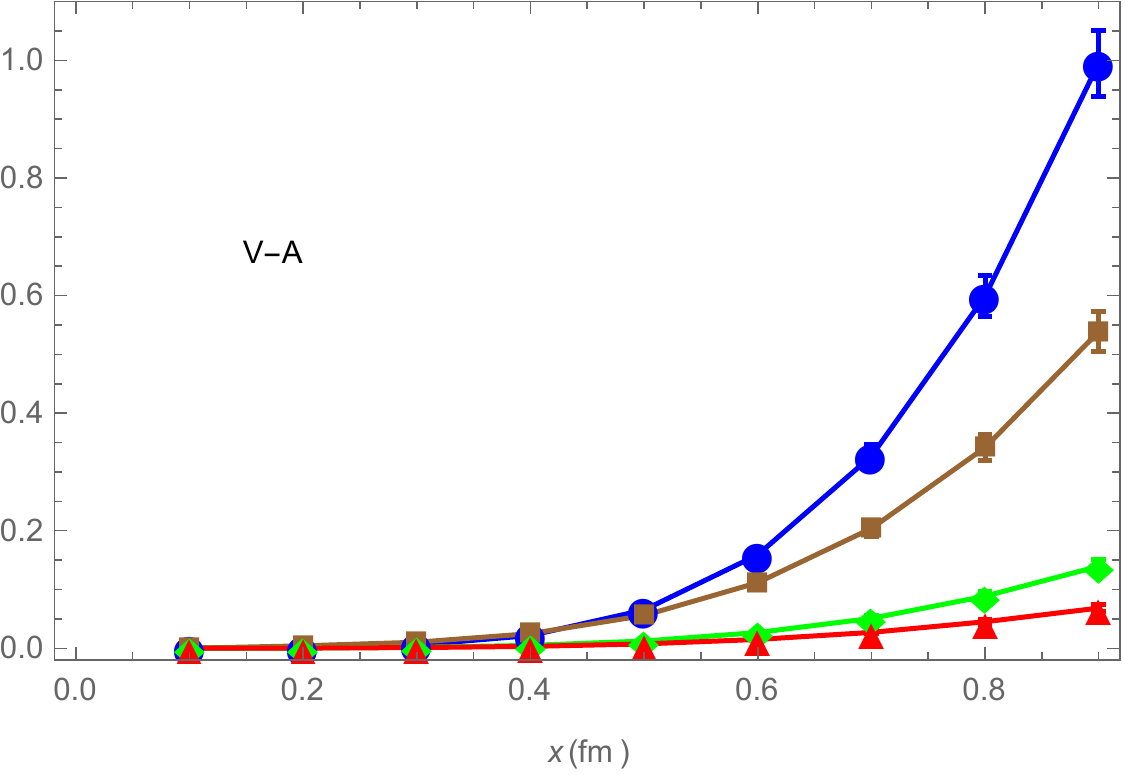}
\includegraphics[width=8cm]{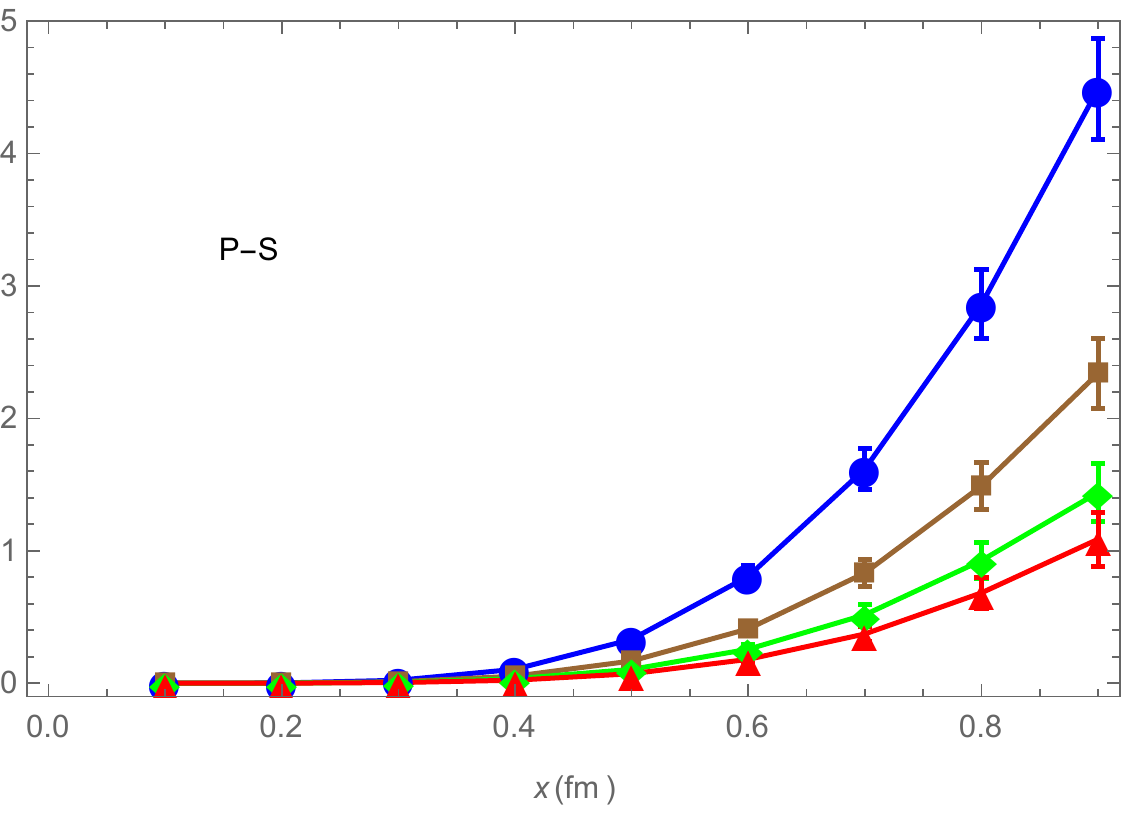}
\caption{ (Color online)   The upper plot is the difference between  the vector and axial correlators $(V(x)-A(x))/(2K_0(x)) $ versus the temporal distance
$x$ (in $fm$). The lower plot is the difference between  the pseudoscalar and scalar correlators  $(P(x)-S(x))/(2K_0(x)) $. $r_{LM} = 0.2 $.
Different curves are for values of the density and holonomy:
$\nu= 1/3$, $n = 0.8$ blue closed circle $\bullet$;
  $\nu= 1/6$, $n =0.8$ brown  box $\Box$;
  $\nu= 1/6$, $n =0.46$ green diamond $\diamond$;
    $\nu= 1/6$, $ n =0.29$ red triangles $\triangle $ .}
\label{fig_diff_density}
\end{center}
\end{figure}

 We  implement only the first two modifications, ignoring the last kinematical one
 and keeping
 (for illustration purposes) the same box size. The results of the calculations with a modified 
 ensembles are shown in Fig. \ref{fig_diff_density}. 
 
 Both the $V-A$ and $P-S$ differences of the correlators decrease,
 as the corresponding modifications are implemented.  As expected, such decrease
 display the tendecy of chiral symmetry breaking effects to ``melt away" at higher $T$.
 
 Furthermore, a careful observer would notice that the decrease in the $(V(x)-A(x))$
 (upper plot)  is much stronger than in the $(P(x)-S(x))$ case (the lower plot). 
 Compare especially the ``highest $T$" points, shown by red triangles.
 
 This means that the restoration of the chiral   $SU(N_f)$  symmetry proceeds 
 more rapidly than the restoration of the chiral $U(1)_a$ symmetry. This is indeed what is expected on general grounds \cite{Shuryak:1993ee}: while the former symmetry 
 is broken spontaneously and gets 
 restored at $T>T_c$, the latter one is broken explicitly by the anomaly and never disappears.
If the instanton-dyon ensembles used would be fully ``unquenched"
from full dynamical simulations including the fermionic determinant, one should see
both phenomena directly. Unfortunately, in this first study we use random
ensembles only, with not-so-small quark mass, and thus full restoration of the chiral   $SU(N_f)$  symmetry, or $V-A=0$ at $T>T_c$, is not there. Yet it is nice to see that
it is at least is getting quite small.

\subsection{Baryonic correlators}
Local currents without derivatives 
with the quantum numbers corresponding to the nucleons (with three flavors, 
the members of the spin-1/2 $SU(3)_f$ octet) and delta resonance (the members of the
spin 3/2 $SU(3)_f$ decuplet) has been defined in \cite{Ioffe:1981kw} and are known as Ioffe currents.

The proton current is
\be J_p= (u^TCd)u-(u^TC\gamma_5d)\gamma_5u
\ee
where the index $T$ means the transposed spinor and $C$ indicates the charge conjugation:
both are needed to write a fermion as an antifermion, to close the bracket (convolute
the color and spinor indices). In such notations the current color and spinor indices 
(not shown) are those of the last quark.

The delta  has a current of a single simple structure, e.g. the charge 3/2 one
\be J_\Delta= (u^TC \gamma_\mu u)u\ee 
and 4 correlators, two ``non-flip" and two "flip" ones.

We had explained the color structure of the correlators, but not yet the spinor one.
Each correlator defined above has two currents which are spinors:
so one can sandwich in any gamma matrix and take the trace. 
Physically,  there are 
 two possible spin structures for the nucleon, with and without a spin flip of the nucleon, corresponding to choices
$ Tr[K], Tr[\gamma_0 K]$. 
 Since one can also study any non-diagonal
correlators, there are in total 6 correlations functions for the nucleon. Two of them
are ``non-flip" and must tend to 1 at small distances, the others go to zero there.
The delta current has only one color structure but more possible spin transitions,
 so in total there are 4 functions.

\begin{figure}[htbp]
\begin{center}
\includegraphics[width=8cm]{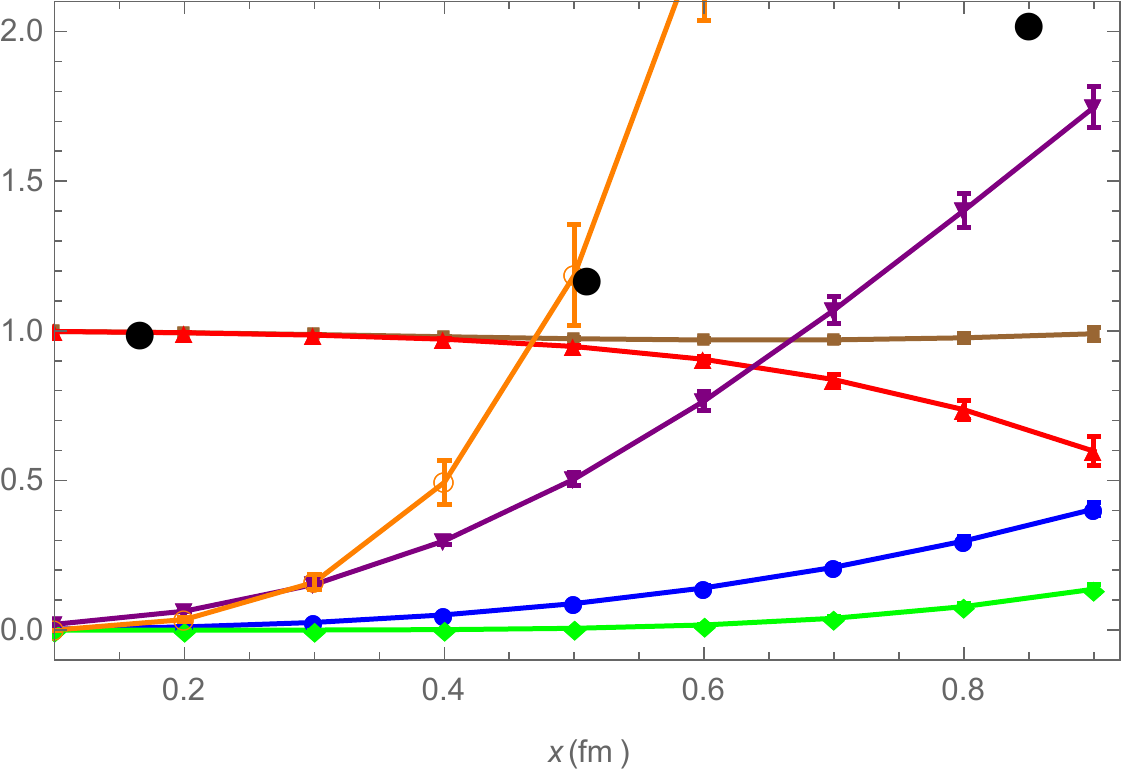}
\caption{ (Color online)  All six Nucleon correlators, see text, versus  the distance $x\, (fm)$, for $r_{12} = 0.2$.  
 The individual black points without line are lattice data from \cite{Chu:1992mn}.
 Those should be compared to the lines with the same symbol.
}
\label{Nucleon_cor}
\end{center}
\end{figure}

\begin{figure}[htbp]
\begin{center}
\includegraphics[width=8cm]{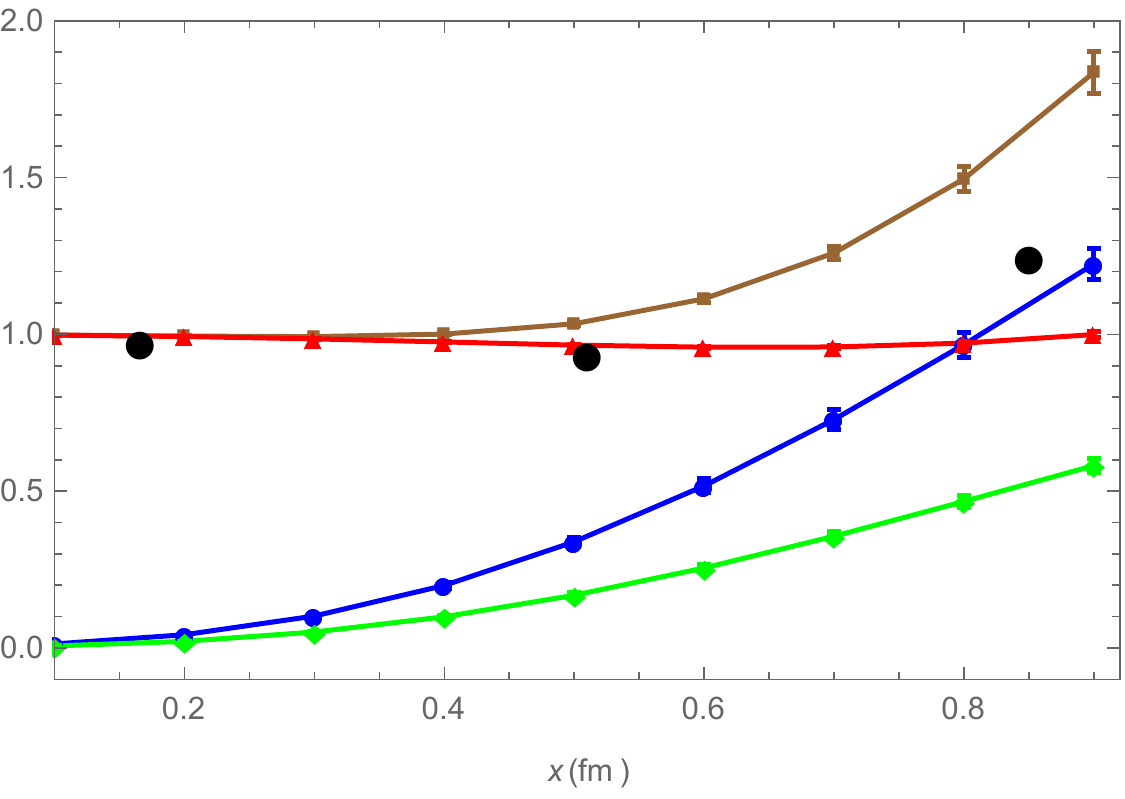}
\caption{ (Color online)  All four delta correlators versus  the distance $x\, (fm)$,  for $r_{12} = 0.2$. 
 The  black closed points without lines are  lattice data \cite{Chu:1992mn}. Those should be compared to the lines with the same symbol.
 }
\label{Delta_cor}
\end{center}
\end{figure}

Results from our simulations for these ten correlation functions are displayed 
in  Fig. \ref{Nucleon_cor} and Fig. \ref{Delta_cor} respectively. 
The normalization is similar to that in the previous subsection, but now to 
the propagation of $three$ free massless quarks $K_0 \sim 1/|x|^9$. Note
that variation of different correlators in this normalization is not very drastic,
although in absolute normalization the correlators would change by about $10^9$
over the range of this plot.

Like for mesonic correlators, we also compare the 
results for some nucleon and delta correlators 
 to the available lattice  data from \cite{Chu:1992mn}, shown by points without connecting lines, and corresponding to quenched quark simulation. 

As we already mentioned,  the main difference 
between the nucleon (spin-1/2 octet baryons) and the Delta (spin-3/2 decuplet baryons)
is due to the fact that the former includes ``good" spin-0 diquark, while the latter
has ``bad" spin-1 diquark. The former one is deeply bound, due mainly to
the operator of the topological  origin, the 't Hooft Lagrangian.

Already in \cite{Shuryak:1993kg} it has been proposed to look at heavy-light
correlation functions, made  of a static quark plus the diquarks. The one of interest
is the $\Lambda$-type
$$ K_\Lambda(x) =< (J_\Lambda(x))^+_n Pexp[\frac{ig }{ 2}\int_0^x A^a_\mu t^a dx^\mu] _{nk} J_\Lambda(0))_k >
$$
including the ``good diquark" current $J_\Lambda(x)_k=u^T_i(x) C \gamma_5 d_j(x) \epsilon_{ijk}$. (For ``bad" diquark the current can be modified by the substitution $\gamma_5\rightarrow \gamma_\mu$.) Note that we have explicitly shown here the color indices, to emphasize the fact that any diquark has spin-color quantum number of an antiquark. In order  to make the correlator gauge invariant one needs to include the connector, the path order exponent, from one point to another. 

Before showing our numerical results for the diquark correlator, two comments are in order.
One is that in a particular
case of $SU(2)$ color group this diquark is a colorless baryon, degenerate with the pion
due to Pauli-Gurcey 
symmetry. While our calculations are for the $SU(3)$, in which no such symmetry is present,
one still may expect certain continuity in $N_c$ and thus a strong attraction in this channel.
The second, following from the first, is that
the ``good" ($ud$ spin-0) diquark is the most attractive channel, thus leading to
phenomenon of color superconductivity at high density. 

\begin{figure}[htbp]
\begin{center}
\includegraphics[width=8cm]{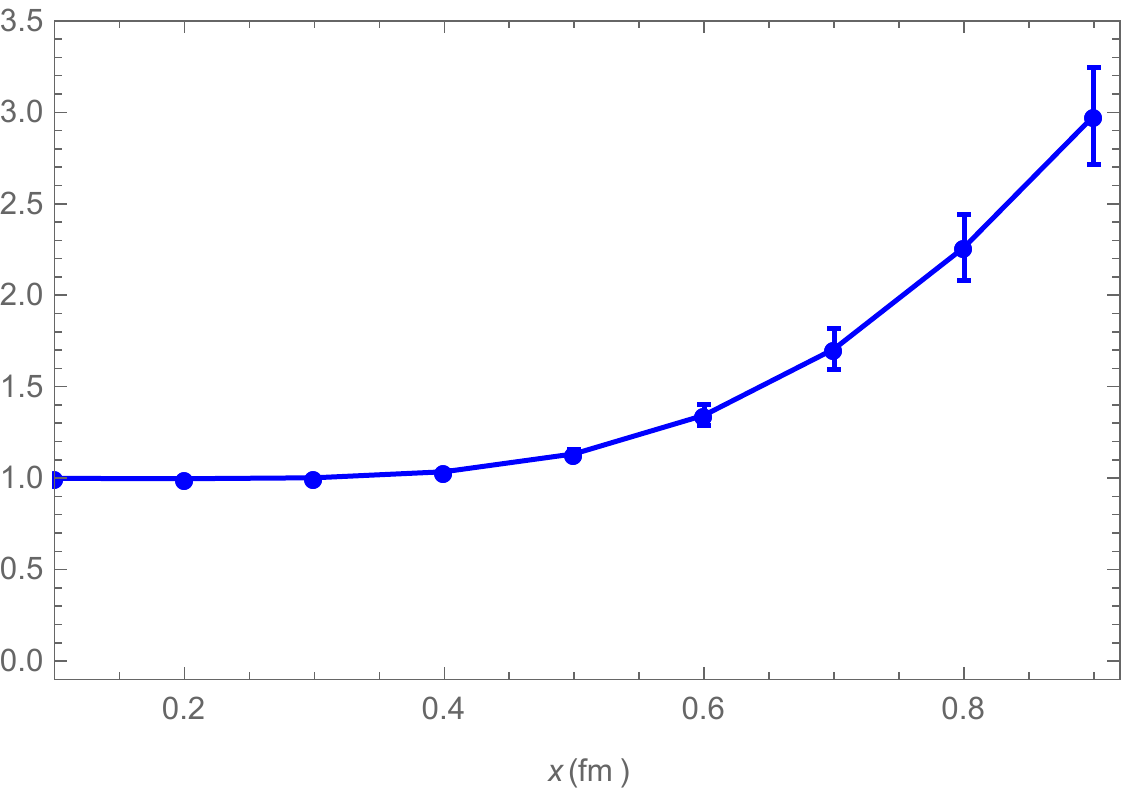}
\caption{ (Color online)  The  normalized correlator for a ``good" ($ud$ spin-0) diquark channel 
versus distance $x\, (fm)$. As other plots, it is done for ensemble of correlated $L-M$ dyons with
 $r_{12} = 0.2$, in  a box of size $2fm$ in time direction and $5fm$ in spatial directions, with 200 zero-modes, which gives a density of $0.8fm^{-4}$.}
\label{Diquark_cor}
\end{center}
\end{figure}

In Fig. \ref{Diquark_cor} we show our measurements of this correlation function. We use an approximation
$Pexp[(ig/2)\int_0^x A^a_\mu t^a dx^\mu] \approx 1$ since its evaluation on our model
 is expensive -- one needs to calculate the gauge fields from all the dyons along the
 straight lone from $x$ to $y$-- and rather unimportant numerically.  
As one can see from this figure, the  normalized correlator goes upward with an increasing distance.
Since the normalization is to the free quark propagation, such behavior indicate attraction
between quarks in this channel.

 Its magnitude is roughly consistent with what was observed in
the instanton liquid model \cite{Schafer:1993ra}. There is a simple 
explanation of its magnitude, based on the analytically known dependence of the 't Hooft Lagrangian
on the number of colors: the factor in the $qq$ channel relative to $\bar q q$ follows from Fiertz
transformation and
 is $f_{N_c}=1/(N_c-1)$. Note that at 
 $N_c\rightarrow \infty $ one has $f_{N_c}\rightarrow 0$. At
 $N_c=2$ one finds  $f_{2}=1$, consistent with Pauli-Gurcey 
symmetry. 
For the case of QCD and our simulations $N_c=3$, thus the relevant factor is  $f_{3}=1/2$.
It is gratifying to see that the simulation results for the diquark show splitting from 1
being indeed  roughly a half of what is observed in the pion channel.


\section{Summary and discussion} \label{sec_summary}

By this paper we started studies of the hadronic correlation functions at nonzero temperatures
$T$, using  numerically generated ensembles of the instanton-dyons. 
Specifically,  in it we addressed the question whether this version of the semiclassical theory, at
its lowest  range of applicability $T\approx T_{min}\sim 100\, MeV$,  can or cannot correctly reproduce many important nonperturbative phenomena in the QCD vacuum.
More specifically, we have done so by explicit evaluation of multiple
mesonic and baryonic two-point correlation functions. 

The 
instanton liquid model 
 has  demonstrated  successful description of those already  in 1990's, and thus one  might naively think that the instanton-dyon ensemble would easily reproduce these functions as well, provided
 the density $n_L$  (per the dyon type)  be in the ballpark of  the instanton density 
 $n_{inst}\sim 1 \, fm^{-4}$. 

What we have found is that this task is by no means  trivial or even simple to fulfill. The reason is the main element of the calculation, the
quark zero modes, are in fact very different for the dyons and instantons.
The dyons are natural at high $T$, at which the temporal  extent of the Matsubara box
$\beta=1/T$ is small, one therefore one can reduce the 4-d theory to its 3-d approximation. The sizes 
of the individual dyons are fixed by the $\beta$ and are small at high $T$. 

However at low $T$ the dyon sizes are getting comparable to the interparticle distances, or even exceed those. If so,  the dyons start to overlap,  partially screening each other. A close pair of $LM$
dyons (in SU(2)) or triplets $L M_1M_2$ (in SU(3)) cancel the fields except in a small
central core. Yet, the topological charge is $not$ screened, and 
 the index theorems thus guarantee  the  existence of quark zero modes. 
As those are localized stronger, to a smaller volume, their normalization 
condition forces them to become locally very strong. As seen in section \ref{sec_zeromodes},
their density grows by about two orders of magnitude.

Since the non-trivial part of the correlation functions of (gauge invariant) currents depend  
on local density of these zero modes, one also observes a very strong dependence of
those on the dyon correlation parameter $r_{LM}$. 

Using accurately known $V-A$ correlator, we were able to tune the value of this parameter
to  $r_{LM}$ to it. After this is done, we calculate several other correlators as well. 

Of particular importance are two strongest effects of the topological origin:
(i) the $P-S$ or $\vec \pi -\vec \delta$ splitting, violating the $U(1)_a$ chiral symmetry, and (ii)
the $ud$ quark pairing into the ``good diquark", present inside the nucleons. 
hadronic phenomenology.

Finally, let us recall that -- both on the lattice and in the semiclassical instanton-dyon theory
-- there remains to study how all the correlation functions and the particular splitting effects
we studied above depend at the temperature. The $V-A$ combination should vanish
for massless quarks at $T>T_c$,  as the $SU(N_f)$ chiral symmetry gets restored. (Or be $O(m)$ if the quark mass is non-zero.)
The other -- and larger -- splitting $\vec \pi -\vec \delta$ is expected to
be nonzero at any $T$, as the $U(1)_a$ chiral symmetry never gets restored. 
While our random ``quenched" ensemble does not fully display this difference,
as we have shown above, it does show it approximately. 




Let us emphasize the importance of high accuracy lattice studies of the issues involved.
This task   has recently been carried out (for vector and axial isovector channels)  
in  Ref.\cite{Tomii:2017cbt}, showing good agreement with  phenomenology and  with the calculation in the framework of the instanton liquid  
model \cite{Schafer:2000rv}. Such studies should be extended to 
the finite temperatures.
While hadrons themselves ``melt" at high temperatures, get large widths and eventually 
completely disappear in hot quark-gluon plasma,
the correlation functions of gauge-invariant operators are well-defined at any temperatures,
and therefore they are the main observables, studied in lattice gauge theory and hadronic phenomenology.

Our paper is based on an ensemble of the instanton-dyons, which has confinement at $T<T_c$.
So we expect the near-realistic spectrum, improved compared to  instanton-based calculations of 1990's. At the other hand, in this
 pilot ``quenched" study, we use ensembles with randomly populated dyons. We expect to
do full dynamical calculations in subsequent works, and see how the
correlators would be affected.

One important issue we would like to understand in those works, by a comparison of the results of this work with
phenomenology, is the issue of ``rigid breaking" of the color group $SU(3)\rightarrow U(1)^2$. The nonzero holonomy
phenomenon, well studied on the lattice, is $<P(T)>\neq 1$. This function provide a $local$ (in $\vec x$)
representation of the $trace$ of the unitary operator, which we parameterize in terms of its eigenvalues $\mu_i$,
which are the main building blocks of the instanton-dyon model. 

The residual local $SU(3)$ rotations do not change the $<P(T)>$ eigenvalues and the the instanton-dyon actions.
Thus they are irrelevant for the non-interacting dyon ensemble we use. 
But they are relevant for the quark propagators used. Indeed, a quark propagating from a dyon located at point $x$
to the dyon located at point $y$ finds two different sets of zero modes at $x$ and $y$, rotated by this residual local $SU(3)$ transformations differently. 
We do not have much first-principle information on the correlation length of these residual local $SU(3)$ rotations,
and therefore approach the issue on try-and-see phenomenological bases. Therefore we will compare
two limiting cases: (i) a {\em ``rigid  breaking"} of $SU(3)\rightarrow U(1)\times  U(1)$", in which the eigenvectors of the Polyakov line have the same direction everywhere, and thus
all dyons have the same color orientation; and (ii)
{\em``random  breaking"}, in which all dyons are  rotated randomly by independent $SU(3)$ matrices.


\end{document}